\documentclass[final,5p,times,twocolumn,authoryear]{elsarticle}

\usepackage[T1]{fontenc}

\usepackage{amssymb}
\usepackage{bm}

\journal{Journal of High Energy Astrophysics}

\begin{document}

\let\phi\varphi
\def\d{\mathrm{d}}
\def\p{\partial}
\def\be{\begin{equation}}
\def\ee{\end{equation}}
\def\bea{\begin{eqnarray}}
\def\eea{\end{eqnarray}}
\def\SdS{Schwarz\-schild--de~Sitter }

\begin{frontmatter}



\title{Possible influence of a cosmic repulsion on large-scale jets: geometric viewpoint}


\author{Petr Slan\'{y}}
\ead{petr.slany@physics.slu.cz}
\affiliation{organization={Research Centre for Theoretical Physics and Astrophysics, Institute of Physics, Silesian University in Opava},
            addressline={Bezru\v{c}ovo n\'{a}m. 13}, 
            city={Opava},
            postcode={CZ-746\,01}, 
            country={Czech Republic}}

\begin{abstract}
Cosmic repulsion represented by a~small positive value of the cosmological constant changes significantly properties of central gravitational fields at large distances, leading to existence of a~static (or turnaround) radius where gravitational attraction of a center is just balanced by cosmic repulsion. Analyzing behavior of radial timelike geodesics in the \SdS spacetime near its static radius we show that the particles with specific energy close to unity have tendency to slow down and cluster just below the static radius, forming clumps which, subsequently, start to expand uniformly due to cosmic repulsion. For central masses of $(10^6$--$10^{11})\,{\rm M_{\odot}}$ and current value of the cosmological constant $1.1\times 10^{-52}$\,${\rm m^{-2}}$, this phenomenon takes place at distances of tens to hundreds of kiloparsecs from the center, being comparable with distances in which huge radio-lobes from some active galaxies were observed.
\end{abstract}




\end{frontmatter}


\section{Introduction}
Outflows of matter are the generic phenomenon connected with various types of astronomical objects ranging from stars to galactic nuclei. Special interest is devoted to collimated outflows--jets. In the case of extragalactic sources, giant jets from some types of Active
Galactic Nuclei (AGNs) belong among the most spectacular phenomena in the Universe. According to the standard picture, jet originates in vicinity of a~super-massive black hole (with several million to several billion solar masses) located at the center of AGN. As a~material in the jet penetrates interstellar medium of the host galaxy and the intergalactic medium beyond, it slows down forming shock front which enables to transform ordered motion of particles to disordered one in the form of an~extended lobe which may be a Mpc or more away \citep{Beg-Bla-Ree:1984:REVMP:,Bla-Mei-Rea:2019:ARASTRA:}. Very impressive, among other things, are the length and stability of jets. Typically, the outer parts of large-scale jets reach kpc-distances but there are some AGNs, whose jets are observed at distances of tens or hundreds of kiloparsecs from the center \citep{Sad-Mor:2002:ASTRJ1:,Eva-etal:2005:MONNR:}. Moreover, AGNs and their relativistic jets are considered as very plausible sources of ultra-high-energy cosmic rays \citep{Rie:2022:UNIVERSE:,Tur-Kol-Stu:2022:SYMMETRY:}. 

Various astronomical observations, including high-redshift supernovae, cosmic microwave background radiation and large-scale structure, indicate current accelerated expansion of our Universe \citep{Rie-etal:1998:ASTRJ1:,Per-etal:1999:ASTRJ2:,Jon-etal:2019:ASTRJ2:,Planck_col:2020:ASTRA:,Boss_col:2015:PHYSR4:}. In the framework of standard relativistic cosmology, it is well described by repulsive effects of empty space or, more generally, so-called \emph{dark energy}, which energy density is characterized by a~small positive value of the cosmological constant $\Lambda\simeq 1.1\times 10^{-52}$\,${\rm m^{-2}}$, corresponding to the energy density $\epsilon_{\Lambda}\simeq 3.3\,{\rm GeV\,m^{-3}}$. 

Existence of a~non-zero, although small, cosmological constant changes significantly also the structure of any central gravitational field which is no longer asymptotically flat but de~Sitter. Such a field can be split into the inner region, in which the gravity of a~central object plays a dominant role, and to the outer region, where the repulsive force of cosmological constant dominates over gravity of the center. The main feature connected with an interplay between gravitational attraction and cosmic repulsion is the existence of a~\emph{static} (or turnaround) \emph{radius} where the both characteristics of the field (attraction and repulsion) are balanced \citep{Stu:1983:BULAI:,Far-Lap-Pra:2015:JCAP:}.

Motions of test particles and photons in asymptotically de~Sitter black-hole spacetimes are analyzed in many papers, including radial, latitudinal or orbital motions, see, e.g., \citet{Stu:1983:BULAI:,Stu-Hle:1999:PHYSR4:,Stu-Hle:2002:ACTPS2:,Stu-Sla:2004:PHYSR4:,Kra:2004:CLAQG:,Stu-Kov:2008:INTJMD:,Ser:2008:PHYSR4:,Hak-etal:2010:PHYSR4:,Char-Stu:2017:EURPJC:,Sla-Stu:2020:EPJC:}. The static radius forms an upper limit on the extension of disk structures around black holes, being, for current value of the cosmological constant and super-massive black holes, comparable with dimensions of large galaxies \citep{Stu-Sla-Hle:2000:ASTRA:,Sla-Stu:2005:CLAQG:,Stu:2005:MODPLA:,Stu-Sla-Kov:2009:CLAQG:,Stu-etal:2020:UNIVERSE:}, and gives also a natural limit on the extension of bound systems in an expanding universe with $\Lambda>0$ \citep{Bus-etal:2003:ASTRJ2:,Stu-Sche:2011:JCAP:,Pav-Ted-Tom:2014:JCAP:,Stu-Hle-Nov:2016:PHYSR4:,Stu-etal:2017:JCAP:,Giu-Far:2019:PHYSDU:}. 

In this paper we restrict our attention to radial motion of massive particles near the static radius of a spherically symmetric \SdS spacetime and show that in the case of central masses $(10^6$--$10^{11})\,{\rm M_{\odot}}$, there are effects which could have some importance for outflows of particles beyond the parent galaxy. The analysis is done for uncharged test particles moving along spacetime geodesics, therefore, our arguments are purely geometric, showing the effect of spacetime itself rather than other (non-gravitational) interactions.

\section{Radial timelike geodesics}
In spherical (Schwarzschild) coordinates $(t,\,r,\,\theta,\,\phi)$ and geometric units $(c=G=1)$, the geometry of the \SdS (SdS) spacetime is given by the line element 
\bea
\d s^2 &=& -(1-2M/r-\Lambda r^2 /3)\d t^2 + \frac{\d r^2}{1-2M/r-\Lambda r^2 /3}  \nonumber \\ 
&+& r^2(\d\vartheta^2 + \sin^2\vartheta\d\phi^2), \label{e1}
\eea
where $M$ is the gravitating mass in the center and $\Lambda>0$ is the cosmological constant.

Radial motion of a~test particle with rest-mass $m$ is described by its 4-velocity $U^{\mu}=\d x^{\mu}/\d\tau=(U^t,\,U^r,\,0,\,0)$, where $\tau$ is particle's proper time, for which $U^{\mu}U_{\mu}=-1$ (in our signature of the metric). Considering a cloud of test particles, their motion, thus, forms a~congruence of radial timelike geodesics. Since the geometry of the SdS spacetime is static, it contains a~timelike (and hypersurface orthogonal) Killing vector field $\xi^{\mu}=\delta^{\mu}_t$ and corresponding projection of particle's 4-velocity $U_t=U_{\mu}\xi^{\mu}\equiv-e$ is a constant of motion. In asymptotically flat Schwarzschild spacetime ($\Lambda=0$), $e$ corresponds to particle's specific energy at radial infinity; in asymptotically de~Sitter SdS spacetime its physical meaning is not so clear but still, in a~good sense, we can interpret it as a~``specific energy'' of a particle. The radial timelike geodesics are, thus, determined by the contravariant components of the 4-velocity
\bea
U^t &=& \frac{e}{1-2M/r-\Lambda r^2 /3} \label{e2} \\
U^r &=& \pm\sqrt{e^2-(1-2M/r-\Lambda r^2 /3)}. \label{e3}
\eea

Radial motion of a~particle with specific energy $e$ is restricted to the region in which
\be\label{e3.1}
e^2\geq V_{\rm eff}(r)\equiv 1-2M/r-\Lambda r^2 /3;
\ee
the auxiliary function $V_{\rm eff}(r)$ plays a~role of the effective potential. 

The gravitational attraction of the center and the cosmic repulsion caused by $\Lambda>0$ are just balanced at the static radius, corresponding to the local (and also global) maximum of the effective potential. For a~particle being at rest at the static radius, $U^{r}=0$ as well as $\d U^{r}/\d\tau=0$; the latter is equivalent to the condition $\d V_{\rm eff}/\d r=0$, giving the location of the static radius at 
\be\label{e4}
r=r_{\rm s}\equiv\sqrt[3]{3M/\Lambda}. 
\ee
The position of such a~particle is, clearly, unstable. However, the static radius is also important for particles which are moving, especially those with the specific energy sufficiently large to cross the radius, but not very large. Lower limit for the specific energy of a~particle crossing the static radius is given by the condition $U^{r}(r_{\rm s})=0$, which means that the turning point of particle's radial motion is just at the static radius. Simple calculation gives the expresion \citep{Stu:1983:BULAI:}
\be\label{e5}
e^2=e^{2}_{\rm s}\equiv 1-3M/r_{\rm s}=1-\sqrt[3]{9\Lambda M^2}.
\ee
We can see that in asymptotically de~Sitter spacetime ($\Lambda>0$), the lower limit of the specific energy for unbound trajectory $e_{\rm s}<1$.

\begin{figure}
\includegraphics[width=.99 \hsize]{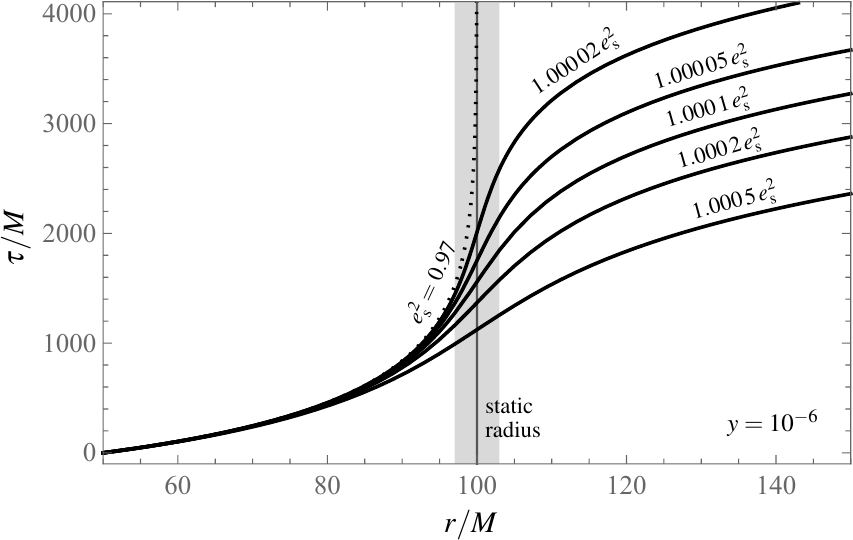}
\caption{Proper time of particles passed from the moment when the particles with prescribed values of the specific energy crossed the radius $r=50M$. The particle with square of its specific energy $e^2=e^2_{\rm s}=0.97$ reaches the static radius in infinite time (dotted curve), while particles with slightly higher values of $e^2$ cross the static radius in finite time. The closer is their specific energy to the limiting value $e_{\rm s}$, the more time they spend in the (shaded) region near the static radius. For illustration we chose the SdS spacetime with so-called cosmological parameter $y=\Lambda M^{2}/3=10^{-6}$.}
\label{f1}
\end{figure}

Now, let's think about particles with specific energies close to but slightly above this limit, i.e. with $e\gtrsim e_{\rm s}$. Such particles, rather than other particles with $e>1$, spend substantial part of their traveling time in the vicinity of the static radius. Fig.~\ref{f1} shows the proper times of particles with various values of the specific energy passed from the moment they crossed the radius $r=50M$. Clearly, the closer is their specific energy to $e_{\rm s}$, the more time they spend near the static radius. 

\begin{figure}
\includegraphics[width=.99 \hsize]{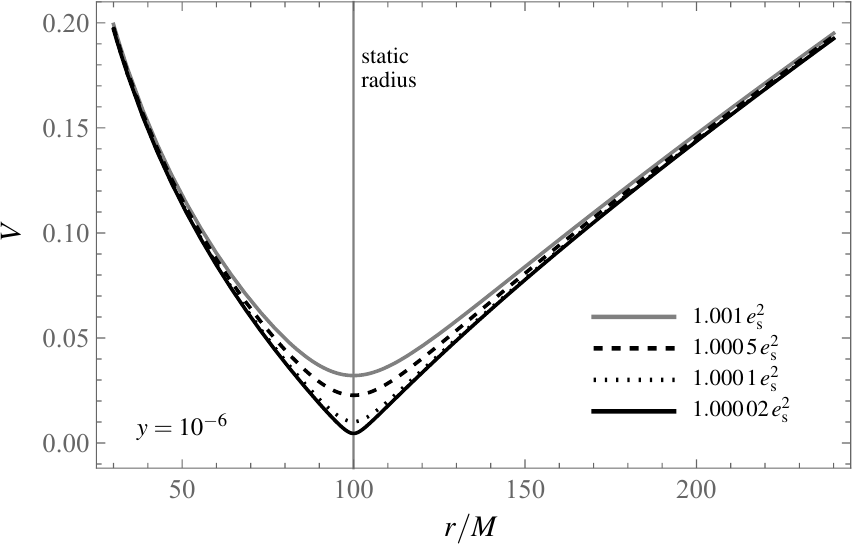}
\caption{Locally measured velocity of particles with specific energy (squared) just above the limiting value $e_{\rm s}^2$. Velocity profiles are presented for particles moving in the SdS spacetime with the cosmological parameter $y=10^{-6}$, in which $r_{\rm s}=100 M$ and $e_{\rm s}^2=0.97$.}
\label{f2}
\end{figure}

Another perspective on the behavior of particles near the static radius is given by the radial dependence of particle's 3-velocity $V$ locally measured by the observer at rest at given radial position. Such a~(static) observer is characterized by the orthonormal tetrad of basis 1-forms:
\bea
\bm{\omega}^{\hat{t}} &=& (1-2M/r-\Lambda r^2 /3)^{1/2}\bm{\d t}, \label{e6}\\
\bm{\omega}^{\hat{r}} &=& (1-2M/r-\Lambda r^2 /3)^{-1/2}\bm{\d r}, \label{e7}\\
\bm{\omega}^{\hat{\vartheta}} &=& r\bm{\d\vartheta}, \label{e8}\\
\bm{\omega}^{\hat{\phi}} &=& r\sin\vartheta\bm{\d\phi}. \label{e9}
\eea
Particle's 3-velocity $V$ with respect to the local static observer is equal to its radial component $V^{\hat{r}}$ given by the relation
\be
V^{\hat{r}}=\frac{U^{\hat{r}}}{U^{\hat{t}}}=\frac{\omega^{\hat{r}}_{\mu}\,U^{\mu}}{\omega^{\hat{t}}_{\mu}\,U^{\mu}}=\sqrt{1-\frac{1-2M/r-\Lambda r^2 /3}{e^2}}. \label{e10}
\ee
Its radial profiles for the values of particle's specific energy just above the limiting value $e_{\rm s}$ are shown in the Fig.~\ref{f2}. Clearly, the particles reach their minimal velocities at the static radius, where the difference between particles with slightly different specific energies is the most evident. Far away from the static radius, the velocity differences of these particles are much smaller. Note that near the static radius these particles move really slowly.

\section{Behavior of radially near particles}
In order to characterize a collective motion of particles with the same specific energy, we analyze covariant divergence of particle's 4-velocity known as the expansion scalar 
\be\label{e11}
\theta=\nabla_{\mu}U^{\mu}=\frac{1}{\sqrt{-g}}\,\p_{\mu}(\sqrt{-g}\,U^{\mu}),
\ee
where $g=\mbox{det}(g_{\mu\nu})$. In the case of radial motion, which is vorticity-free as well as shear-free, the expansion scalar tells us how the radial distance of near particles, located at radial positions $r$ and $r+\delta r$, changes with time. When $\theta>0$, the particles recede from each other, when $\theta<0$, the particles approach each other. For radially outgoing particles, the expansion scalar is given by the expression
\be\label{e12}
\theta(r;\,e^2)=\frac{2(e^2-1)+3M/r+\Lambda r^2}{r\sqrt{e^2-(1-2M/r-\Lambda r^2 /3)}}.
\ee

\begin{figure}
\includegraphics[width=.99 \hsize]{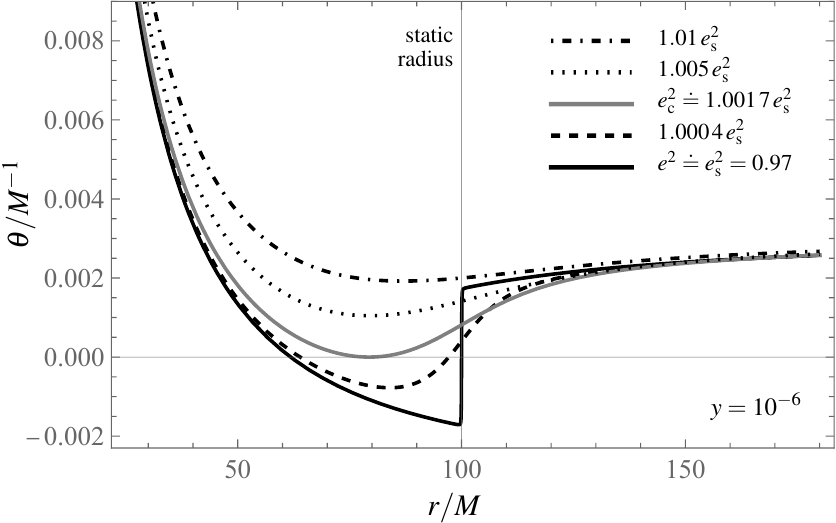}
\caption{Radial profiles of the expansion scalar characterizing radial motions of particles with specific energies close to but above the critical value $e_{\rm s}$ in the SdS spacetime with the cosmological parameter $y=10^{-6}$. Values of the specific energy squared for a~given profile are specified in the legend.}
\label{f3}
\end{figure}

Again we restrict our attention on particles which are able to cross the static radius, i.e., with specific energies above the critical value $e_{\rm s}$. In asymptotically flat Schwarzschild spacetime ($\Lambda=0$), the expansion scalar $\theta(r)$ is monotonically decreasing function of $r$ independently of the value $e>1$. On the other hand, in asymptotically de~Sitter SdS spacetime, for particles with specific energies $e_{\rm s}<e<1$, there is a local minimum, which for specific energies just above the critical value $e_{\rm s}$ is located even under the static radius. Moreover, for particles with specific energies in a~small interval $e_{\rm s}<e<e_{\rm c}$,
where
\be\label{e13}
e^{2}_{\rm c}=1-3\sqrt[3]{9\Lambda M^{2}/32},
\ee
the local minimum is negative, indicating convergence rather than divergence of these particles near (and under) the static radius. Radial profiles of the expansion scalar for various particles' specific energies close to but above the critical value $e_{\rm s}$ are plotted in Fig.~\ref{f3}. Although the range of energies corresponding to radially converging particles is rather small, the region, in which these particles approach each other, $r_{\rm c}<r<r_{\rm s}$, where
\be\label{e14}
r_{\rm c}=\frac{1}{2}(\sqrt{5}-1)\,r_{\rm s}\doteq 0.62\, r_{\rm s},
\ee
is sufficiently large.

\section{Conclusions}
Cosmic repulsion characterized by a~small positive value of the cosmological constant changes significantly properties of central gravitational fields at large distances, leading to existence of the static radius where gravitational attraction of a center is just balanced by cosmic repulsion. 
Current value of the cosmological constant and, therefore, also related position of the static radius are given by current values of the dark energy density parameter $\Omega_{\Lambda}$ and the Hubble parameter $H_{0}=100 h\,{\rm km\,s^{-1}\,Mpc^{-1}}$ (so-called Hubble constant) through relations
\bea\label{e15}
\Lambda &=& 3\Omega_{\Lambda}H_{0}^{2}/c^2=3.5\times 10^{-52}\,\Omega_{\Lambda}h^2\,{\rm m^{-2}}, \\
r_{\rm s} &=& 75.5\,(\Omega_{\Lambda}h^2)^{-1/3}(M/M_{\odot})^{1/3}\,{\rm pc}. \label{e16}
\eea
Further we take $\Omega_{\Lambda}=0.689$, $h=0.677$, inferred for the spatially flat $\Lambda$CDM model from \emph{Planck} CMB power spectra, in combination with CMB lensing reconstruction and BAO \citep{Planck_col:2020:ASTRA:}, which give the value $\Lambda\doteq 1.1\times 10^{-52}$\,${\rm m^{-2}}$. Note that there is some discrepancy between values of the Hubble constant determined from \emph{Planck} CMB power spectra and by other methods based on the distance-ladder techniques, as these methods give somewhat higher values of $H_{0}$, see, e.g., \cite{Rie-etal:2018:ASTRJ2:,Fre-etal:2019:ASTRJ2:,Tau-etal:2019:ASTRA:} and references therein.

For $\Lambda\doteq 1.1\times 10^{-52}$\,${\rm m^{-2}}$, the static radii for various central masses are given in Table~\ref{t1}.
Clearly, in the case of super-massive black holes or even masses corresponding to the whole galaxies, the static radii are located at distances, in which large-scale jets and radio lobes from some AGNs are observed.

Presented analysis of radial timelike geodesics of the \SdS spacetime indicates possible, although unexpected, influence of the cosmic repulsion on outflows of matter from gravitationally bound systems like galaxies, for example on large-scale jets from AGNs. The main effect resides in the tendency of particles with specific energy close to unity to slow down and cluster near the static radius, forming clumps which, subsequently, start to expand and accelerate almost uniformly due to cosmic repulsion. 

\begin{table}%
\begin{tabular}{lcccccccc}\hline
$M/M_{\odot}$ & 10 & $10^2$ & $10^6$ & $10^8$ & $10^{9}$ & $10^{10}$ & $10^{11}$ \\ 
$r_{\rm s}$/kpc & 0.24 & 0.52 & 11 & 52 & 110 & 240 & 520 \\ \hline
\end{tabular}
\caption{Positions of the static radius $r_{\rm s}$ in kiloparsecs for given central mass $M$ (in units of solar mass) and current values $\Omega_{\Lambda}\doteq 0.689$, $h=0.677$.}
\label{t1}
\end{table}

\section*{Acknowledgements}
The work was done as a part of the project SGS/24/2024 financed by the Silesian University in Opava, Czech Republic.
Support from the Research Centre for Theoretical Physics and Astrophysics, being a part of the Institute of Physics of the Silesian University in Opava, Czech Republic, is also greatly acknowledged.








\end{document}